\documentclass[aps,pre,twocolumn,showpacs,superscriptaddress,groupedaddress,acknowledgment]{revtex4-1}  % for review and submission
\usepackage{dcolumn}   % needed for some tables
\usepackage{bm}        % for math

\usepackage{amsmath,comment}
\usepackage{braket}
\usepackage{amsfonts}
\usepackage{graphicx}
\usepackage{amsthm}
\usepackage{color}
\usepackage{setspace}
\usepackage{amssymb}
\usepackage{latexsym}
\usepackage{here}

\theoremstyle{definition}

\newtheorem*{theorem*}{Theorem}

\newtheorem*{definition*}{Definition}

\begin{document}
\title{Generalized Gibbs ensemble in a nonintegrable system \\with an extensive number of local symmetries}
\author{Ryusuke Hamazaki}
\affiliation{Department of Physics, University of Tokyo, Bunkyo-ku, Tokyo 113-0033, Japan}
\author{Tatsuhiko N. Ikeda}
\affiliation{Department of Physics, University of Tokyo, Bunkyo-ku, Tokyo 113-0033, Japan}
\affiliation{Department of Physics, Harvard University, Cambridge, Massachusetts 02138, USA}
\author{Masahito Ueda}
\affiliation{Department of Physics, University of Tokyo, Bunkyo-ku, Tokyo 113-0033, Japan}
\affiliation{RIKEN Center for Emergent Matter Science (CEMS), Wako, Saitama 351-0198, Japan}

\date{\today}
\begin{abstract}
We numerically study the unitary time evolution of a nonintegrable model of hard-core bosons with an extensive number of local $\mathbb{Z}_2$ symmetries.
We find that the expectation values of local observables in the stationary state are described better by the generalized Gibbs ensemble (GGE) than by the canonical ensemble.
We also find that the eigenstate thermalization hypothesis fails for the entire spectrum, but holds true within each symmetry sector,
which justifies the GGE.
In contrast, if the model has only one  global $\mathbb{Z}_2$ symmetry or a size-independent number of local $\mathbb{Z}_2$ symmetries, we find that the stationary state is described by the canonical ensemble.
Thus, the GGE is necessary to describe the stationary state even in a nonintegrable system if it has an extensive number of local symmetries.
\end{abstract}
\maketitle
\section{Introduction}
Conserved quantities play a crucial role in characterizing stationary states in isolated quantum systems \cite{Polkovnikov11,Nandkishore14,Eisert15,Gogolin15,DAlessio15}.
When the total energy is the only conserved quantity,
the stationary state is expected to be described by the (micro)canonical ensemble \cite{Neumann29,Tasaki98,Goldstein06,Popescu06,Sugita06,Reimann07,Rigol08,Reimann08,Reimann10,Linden09,Goldstein10,Sugiura12,Sugiura13,Reimann15,Goldstein15,Trotzky11}.
The eigenstate thermalization hypothesis (ETH) is a likely candidate for explaining the validity of the canonical ensemble in nonintegrable systems \cite{Jensen85,Deutsch91,Srednicki94,Rigol08,Santos10a,Beugeling14,Kim14,Khodja15,Reimann15E,Fratus15}.
In contrast, in integrable systems \cite{Rigol07,Rigol09,Kaminishi15E,Kinoshita06,Gring12,Langen13,Langen15} or systems showing many-body localization \cite{Basko06,Pal10,Gogolin11,Iyer13,Mondaini15,Ponte15,Tang15,Schreiber15,Smith15}, the stationary state cannot be described by the canonical ensemble due to nontrivial conserved quantities.

The generalized Gibbs ensemble (GGE) successfully describes stationary states
in integrable systems whose Hamiltonian can be mapped to a quadratic form that describes quasiparticles \cite{Rigol07,Cazalilla06,Kollar08,Rigol09,Cassidy11,Calabrese11,Cazalilla12,Fagotti13,Langen15}.
The GGE is constructed in terms of the numbers of quasiparticles in each mode, $\hat{n}_\alpha$, and given by $\hat{\rho}_{\rm GGE}=e^{-\sum_\alpha \lambda_\alpha \hat{n}_\alpha}/Z_{\rm GGE}$.
Here $Z_{\rm GGE}\equiv \text{Tr}[e^{-\sum_\alpha \lambda_\alpha \hat{n}_\alpha}]$
and the parameters $\lambda_\alpha$ are determined from the initial values of $\hat{n}_\alpha$.
The GGE has also been applied \cite{Mossel12,Caux12,Pozsgay13,Fagotti13SB,Kormos13,Mussardo13,Pozsgay14CQ,Pozsgay14,Pozsgay14QQ,Wouters14,Essler15,Ilievski15,Zill15}
to the Bethe-ansatz-solvable systems \cite{Sato12,Ikeda13,Kaminishi15R,Deguchi15,Alba15}.
These integrable systems have sufficiently many conserved quantities so that each energy eigenstate can be identified.
This feature is also seen in systems exhibiting strong many-body localization,
where the GGE is expected to be constructed from the local integrals of motion \cite{Vosk13,Serbyn13,Huse14}.

Thus, for a comprehensive understanding of the stationary states, it is of interest to study models with moderate numbers of conserved quantities.
The stationary state is described by the canonical ensemble if the total energy is the only conserved quantity.
On the other hand, when there are sufficiently many conserved quantities to identify eigenstates, the GGE is necessary.
Then, the following question arises: how many conserved quantities are required for the GGE to be needed to describe the stationary state?

In this paper, we show that the GGE is necessary to describe stationary states even in a nonintegrable system if it has an extensive number of local symmetries.
We numerically study a nonintegrable model of hard-core bosons with an extensive number of local $\mathbb{Z}_2$ symmetries that lead to many conservation laws.
We show that the expectation values of local observables in the stationary states are described by the GGE rather than the canonical ensemble.
We argue that this is because the ETH holds true not for the entire spectrum but for each symmetry sector.
For the sake of comparison, we examine a model that involves only one global $\mathbb{Z}_2$ symmetry or a size-independent number of local $\mathbb{Z}_2$ symmetries, and show that the canonical ensemble works and that the GGE is not necessary for these models.

The rest of this paper is organized as follows. 
In Sec. II, we define a model with an extensive number of local symmetries. 
In Sec. III, we analyze unitary time evolutions starting from two distinct initial states.
We argue that the stationary state is described by the GGE rather than the canonical ensemble.
In Sec. IV, we confirm the results obtained in Sec. III by varying the system size.
In Sec. V, we show that the ETH fails for the entire spectrum, but holds true for each symmetry sector.
In Sec. VI, we study models with fewer than extensive local symmetries, and show that the canonical ensemble works in this case and that the GGE is not necessary.
In Sec. VII, we summarize the main results of this paper and discuss some future directions.
Some explanatory or supplemental materials are relegated to appendices to avoid digressing from the main subjects.

\section{A model with an extensive number of local symmetries}\label{defmodel}
\begin{figure}[tbp]
\begin{center}
\includegraphics[width=\linewidth]{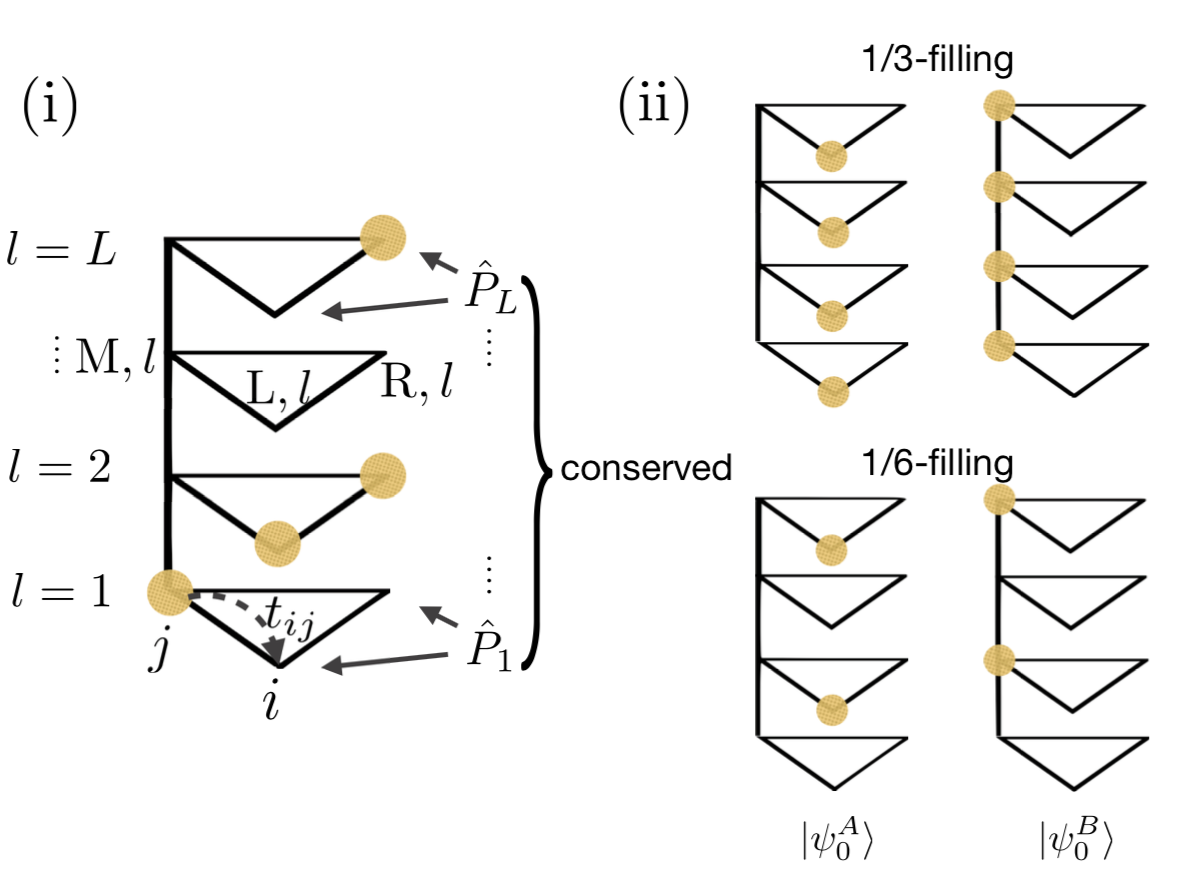}
\caption{(i) Our model of hard-core bosons with $L=N_b=N_s/3=4$ (1/3-filling). 
Bosons can hop between connected sites with hopping energy $t_{ij}$.
Each layer labeled by $l\:(1\leq l \leq L)$ has a local $\mathbb{Z}_2$ symmetry with respect to the swap of the sites L and R.
The swapping operator at the $l$th layer is denoted by $\hat{P}_l$.
(ii) Two initial states $\ket{\psi^A_0}$ (left) and $\ket{\psi^B_0}$ (right), where bosons are placed at $(\mathrm{L},l)$ and $(\mathrm{M},l)$, respectively. For a 1/3-filling (up), every layer is occupied by one boson, and for a 1/6-filling (down), every even layer is occupied by one boson.
}
\label{bose}
\end{center}
\end{figure}

We consider a nonintegrable model of $N_b$ hard-core bosons distributed over $N_s$ sites that are arranged in $L$ layered triangles $(N_s=3L)$ as illustrated in Fig. \ref{bose} (i).
We label each site $i\:(1\leq i\leq N_s)$ by two indices $(s,l)$, where
$l\:(=1,2,\cdots, L)$ labels the layer and $s\:(=\mathrm{L,M,R})$ labels the location in each layer. 

The Hamiltonian is 
\begin{equation}\label{Ham}
\hat{H}=-\sum_{\langle i,j \rangle}t_{ij}(\hat{b}^\dag_i \hat{b}_j +\text{H.c.})\: .
\end{equation} 
Here $\hat{b}_i$ is an annihilation operator of a hard-core boson at a site $i$, $t_{ij}\in\mathbb{R}$ is a hopping energy,
and $\langle i,j\rangle$ denotes a pair of neighboring sites $(i<j)$.
%In this paper, we consider the case in which the total number of hard-core bosons is $L$, namely, the filling is 1/3.
%We have confirmed that the qualitative conclusions of this paper remain valid for other fillings.

We assume that the hopping energy $t_{ij}$ satisfies
\begin{equation}\label{symcond}
t_{\text{M}l,\text{L}l}=t_{\text{M}l,\text{R}l}\: ,
\end{equation}
which guarantees a local $\mathbb{Z}_2$ symmetry associated with the swapping operator $\hat{P}_l\:(1\leq l \leq L)$ for each layer.
This operator swaps the sites (L,$l$) and (R,$l$) (see Fig. \ref{bose} (i)), and satisfies $\hat{P}_l\hat{b}_{\text{L}l}\hat{P}^\dag_l=\hat{b}_{\text{R}l}$ and $\hat{P}_l\hat{b}_{\text{R}l}\hat{P}^\dag_l=\hat{b}_{\text{L}l}$.
We can write $\hat{P}_l$ as 
$
\hat{P_l}= \hat{I} +\hat{b}_{\text{L}l}^\dag \hat{b}_{\text{R}l}
+\hat{b}_{\text{R}l}^\dag \hat{b}_{\text{L}l} - (\hat{b}_{\text{L}l}^\dag \hat{b}_{\text{L}l}-\hat{b}_{\text{R}l}^\dag \hat{b}_{\text{R}l})^2
$, which satisfies $[\hat{H},\hat{P}_l]=0$ and $\hat{P}_l^2=1$.
The eigenvalues of $\hat{P}_l$ are $q_l=\pm 1$, which we call the positive and negative $\mathbb{Z}_2$ parities.
By mapping the hard-core bosons to the spin 1/2 operators, we can show that
 $\hat{P}_l$ works as the projection operator onto the spin singlet $(q_l=-1)$ and triplet $(q_l=+1)$ states formed by the spins on (L,$l$) and (R,$l$).

The system thus has a symmetry group that is given by $G=\bigotimes_{l=1}^L \mathbb{Z}_2$.
Since $G$ is abelian, the energy eigenstates are divided into the $|G|=2^L$ symmetry sectors \cite{Georgi}, which are characterized by a set of $\mathbb{Z}_2$ parities $\mathbf{q}\equiv(q_l)_{l=1}^L$.
If we label the symmetry sectors by $\mathbf{q}$, the entire Hilbert space $\mathcal{H}$ of the system is divided into $\mathcal{H}=\bigoplus_{\mathbf{q}}\mathcal{H}_\mathbf{q}$.

To remove unwanted symmetries and accidental degeneracies, we add randomness to $t_{ij}$
by setting $t_{\mathrm{M}l,\mathrm{L}l}$(=$t_{\mathrm{M}l,\mathrm{R}l}$), $t_{\mathrm{L}l,\mathrm{R}l}$, and $t_{\mathrm{L}l,\mathrm{R}(l+1)}$ as 
\begin{align}\label{rand}
t_{ij}={t_\mathrm{hop}}(1+\epsilon_{ij})\:, 
\end{align}
where $\epsilon_{ij} \in [-0.5,0.5]$ is randomly chosen according to the uniform measure.
This randomness removes all degeneracies and most of the symmetries except for the $\mathbb{Z}_2$ symmetry.
We note that the eigenenergy spacings obey the Wigner-Dyson statistics within each parity sector $\mathcal{H}_\mathbf{q}$ that contains sufficiently many eigenstates (see Appendix \ref{level}).

\section{Long-time evolutions from two initial states}\label{quench}
We consider two initial states $\ket{\psi_0}=\ket{\psi^A_0}$ and $\ket{\psi^B_0}$, where bosons are placed at $(\mathrm{L},l)$ and $(\mathrm{M},l)$, respectively [Fig. \ref{bose} (ii)].
The time evolutions from these initial states will be referred to as case A and case B.
We consider the cases of 1/3-filling, where $N_b=L$ and one boson is placed at every layer ($l=1,2,\cdots,L$),  and 1/6-filling, where $N_b=L/2$ and one boson is placed at every even layer ($l=2,4,\cdots,L$).
While $\ket{\psi^A_0}$ extends over different $\mathcal{H}_\mathbf{q}$'s, 
$\ket{\psi^B_0}$ belongs to a single sector $\mathcal{H}_{\mathbf{q}_1}$, where $\mathbf{q}_1=(+1,+1,\cdots,+1)$ (see Appendix \ref{proj}). 
Both of the initial states have the total conserved energy $\braket{\psi^A_0|\hat{H}|\psi^A_0}=\braket{\psi^B_0|\hat{H}|\psi^B_0}=0$, which corresponds to the infinite temperature (see Appendix \ref{inf}).

The state at time $t$ is given by $\ket{\psi(t)}=e^{-\frac{i\hat{H}t}{\hbar}}\ket{\psi_0} =\sum_\alpha c_\alpha e^{-\frac{i{E}_\alpha t}{\hbar}}\ket{E_\alpha}$.
Here $\ket{E_\alpha}$ is an energy eigenstate with eigenenergy $E_\alpha$, and $c_\alpha = \braket{E_\alpha|\psi_0}$.

The long-time average of a local observable ${\hat{\mathcal{O}}}$ is described by the diagonal ensemble if there are no degeneracies among the eigenstates \cite{Linden09,Rigol08}:
\begin{align}\label{diag}
\lim_{T\rightarrow \infty}\frac{1}{T}\int_0^T{\braket{\psi(t)|{\hat{\mathcal{O}}}|\psi(t)}} = \mathrm{Tr}[ \hat{\rho}_\mathrm{d} {\hat{\mathcal{O}}} ]\:,
\end{align}
where $\hat{\rho}_\mathrm{d}\equiv\sum_\alpha |c_\alpha|^2 \ket{E_\alpha}\bra{E_\alpha} $.
When a large number of eigenstates are superposed in the initial states, temporal deviations from the prediction of the diagonal ensemble become sufficiently small \cite{Reimann08,Linden09,Short11,Reimann12,Short12}.
Note that the diagonal ensemble has an exponentially large number of microscopic parameters $|c_\alpha|^2$.

We define the canonical ensemble and the GGE which will be used to describe the stationary state with a few parameters.
The canonical ensemble is defined as
\begin{align}\label{canonicale}
\hat{\rho}_\mathrm{can} = \frac{1}{Z_\mathrm{can}} e^{-\beta \hat{H}}\:,
\end{align}
where $Z_\mathrm{can} =\mathrm{Tr}[e^{-\beta \hat{H}}]$.
Here the inverse temperature $\beta$ is uniquely determined from the total energy
$
E_0 \equiv \braket{\psi_0|\hat{H}|\psi_0}=\mathrm{Tr}[\hat{\rho}_\mathrm{can}\hat{H}].
$
On the other hand, the GGE in our system is constructed as
\begin{align}\label{GGE}
\hat{\rho}_\mathrm{GGE} &= \frac{1}{Z_\mathrm{GGE}} e^{-\tilde{\beta }\hat{H}-\sum_{l=1}^L \lambda_l \hat{P}_l}\:,
\end{align}
where $Z_\mathrm{GGE} =\mathrm{Tr}[e^{-\tilde{\beta }\hat{H}-\sum_{l=1}^L \lambda_l \hat{P}_l}]$.
Here $\tilde{\beta}$ and $\lambda_l\:(1\leq l\leq L)$ are uniquely determined from the conditions 
$
\braket{\psi_0|\hat{H}|\psi_0}=\mathrm{Tr}[\hat{\rho}_\mathrm{GGE}\hat{H}]
$ and
$
\braket{\psi_0|\hat{P}_l|\psi_0}=\mathrm{Tr}[\hat{\rho}_\mathrm{GGE}\hat{P_l}]\:(1\leq l \leq L)
$.

\begin{figure}[tbp]
\begin{center}
\includegraphics[width=\linewidth]{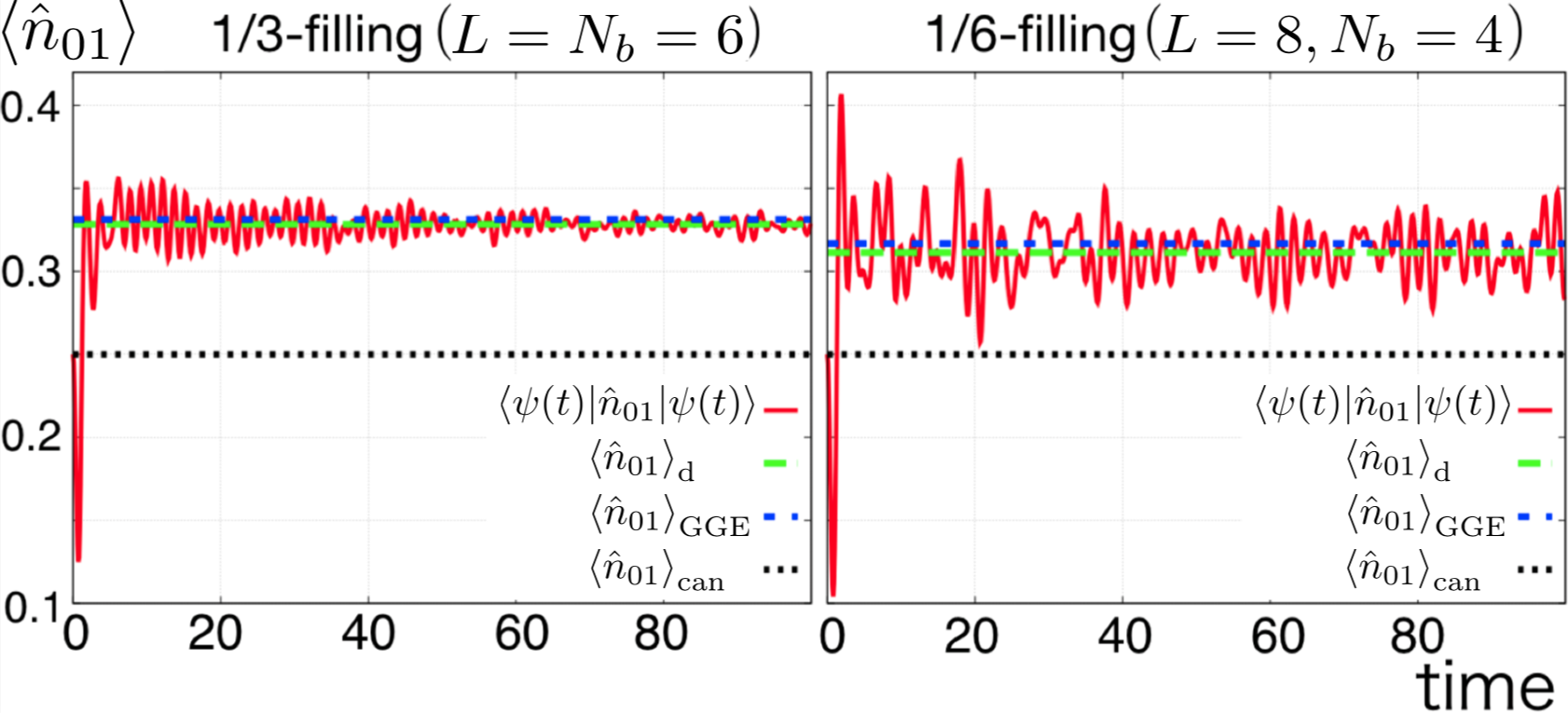}
\caption{(Color Online) Typical time evolutions of the expectation value of $\hat{n}_{01}$ for case A (solid curve). The time is measured in units of $\hbar/t_\mathrm{hop}$.
The left and right panels show the result for the 1/3-filling ($L=N_b=6$) and that for the 1/6-filling ($L=8, N_b=4$), respectively.
The stationary state is well described by the diagonal ensemble (yellow long dashed line).
While the GGE (blue short dashed line) also describes the stationary state well, the canonical ensemble (black dotted line) does not (see Appendix \ref{inf} for details).
}
\label{timeevolve}
\end{center}
\end{figure}

As observables, we take the (normalized) number of hard-core bosons with a given momentum $\mathbf{k}=(k_x,k_y,0)$.
Its average along the $z$ (vertical) direction gives
$\hat{n}(k_x,k_y)=\frac{1}{2^2N_b}\sum_{{i,j}}\delta_{z_i,z_j}e^{-i\mathbf{k}\cdot (\mathbf{r}_i-\mathbf{r}_j)}\hat{b}^\dag_i \hat{b}_j$.
Here $\mathbf{r}_i=(x_i,y_i,z_i)$ denotes the coordinate of the site $i$ (the lattice constant is set to unity).
Specifically, we consider $\hat{n}_{00}\equiv \hat{n}(0,0),{n}_{01}\equiv \hat{n}(0,\pi)$, and $\hat{n}_{11}\equiv \hat{n}(\pi,\pi)$ in the following discussions.

Figure \ref{timeevolve} demonstrates typical time evolutions of the expectation value of $\hat{n}_{01}$ for case A.
The left and right panels show the result of the 1/3-filling ($L=N_b=6$) and that of the 1/6-filling ($L=8, N_b=4$), respectively.
The predictions of the diagonal ensemble, the canonical ensemble, and the GGE, which are respectively given by $\braket{\hat{n}_{01}}_\mathrm{d}\equiv \mathrm{Tr}[\hat{\rho}_\mathrm{d}\hat{n}_{01}]$, $\braket{\hat{n}_{01}}_\mathrm{can}\equiv \mathrm{Tr}[\hat{\rho}_\mathrm{can}\hat{n}_{01}]$, and $\braket{\hat{n}_{01}}_\mathrm{GGE}\equiv \mathrm{Tr}[\hat{\rho}_\mathrm{GGE}\hat{n}_{01}]$, respectively, are also shown.
The expectation value relaxes to the prediction of the diagonal ensemble for large $t$ with small temporal fluctuations.
We find that the GGE describes the stationary state and the diagonal ensemble very well, whereas the canonical ensemble does not.
This result highlights our key finding regardless of the value of the filling: the GGE is necessary to describe the stationary state in a nonintegrable system with an extensive number of local symmetries.
In the next section, we confirm this observation in more detail by focusing on the case of 1/3-filling ($L=N_b$).

\section{Validity of the Generalized Gibbs Ensemble: Scaling Results}\label{atagge}
By varying the system size, we quantitatively analyze how well the GGE describes the stationary state compared with the canonical ensemble.
We define the relative difference between the canonical ensemble and the diagonal ensemble, and a similar quantity for the GGE as follows:
\begin{align}\label{rela}
\overline{\delta n_\mathrm{can}}&\equiv\overline{\left|\frac{\braket{\hat{n}}_\mathrm{d}-\braket{\hat{n}}_\mathrm{can}}{{\braket{\hat{n}}_\mathrm{d}}}\right|}, \nonumber \\
\overline{\delta n_\mathrm{GGE}}&\equiv\overline{\left|\frac{\braket{\hat{n}}_\mathrm{d}-\braket{\hat{n}}_\mathrm{GGE}}{\braket{\hat{n}}_\mathrm{d}}\right|}\:.
\end{align}
Here $hat{n}$ represents $\hat{n}_{00},\hat{n}_{01}$ or $\hat{n}_{11}$, and $\overline{\cdots}$ denotes the average over 20 sample Hamiltonians having different randomness in $t_{ij}$ [see Eq. (\ref{rand})].

\begin{figure}[tbp]
\begin{center}
\includegraphics[width=\linewidth]{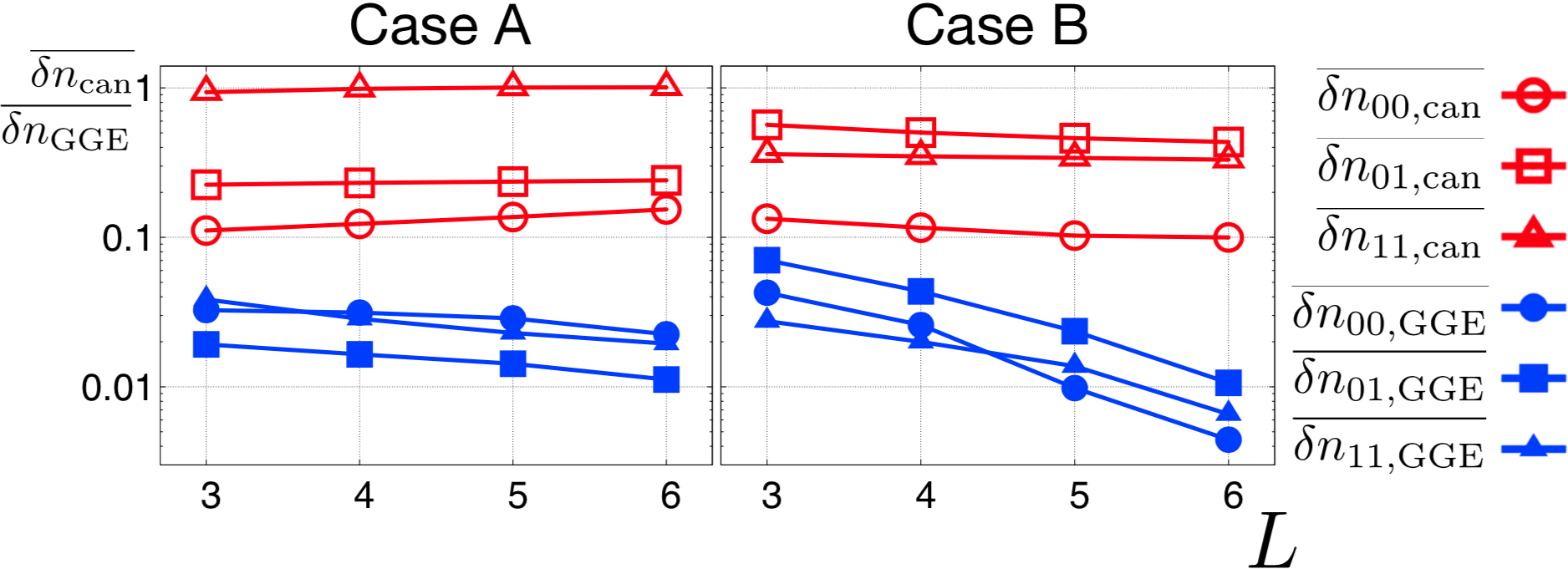}
\caption{(Color Online) Relative differences of the canonical ensemble (open) and the GGE (filled) compared with the diagonal ensemble [see Eq. (\ref{rela})]  for $\hat{n}_{00}\text{ (circle)},\hat{n}_{01}\text{ (square)}$, and $\hat{n}_{11}\text{ (triangle)}$. The left and right panels show the results for case A and case B, respectively.
For both of the initial states, the relative difference for the canonical ensemble does not appear to decrease with increasing $L$, whereas it does for the GGE.}
\label{diff}
\end{center}
\end{figure}

Figure \ref{diff} shows that the relative difference of the GGE is about ten times smaller than that of the canonical ensemble.
We note that the relative difference stays more than 10\% for the canonical ensemble, whereas 
it tends to decrease with increasing the system size for the GGE.

Figure \ref{diff} also shows some distinction between case A and case B, concerning the $L$-dependence of the relative difference of the GGE.
The relative difference decreases less rapidly in case A than in case B with increasing $L$. 
This is due to the mixing of the symmetry sectors with negative parity in case A, as detailed in the next section.

\section{Verification of the ETH for each symmetry sector}\label{RETH}
In this section, we investigate the ETH to understand why the GGE works for our model, whereas the canonical ensemble does not.
The ETH is a statement for  the eigenstate expectation value (EEV) of a local observable ${\hat{\mathcal{O}}}$, i.e. $\braket{E_\alpha|\hat{\mathcal{O}}|E_\alpha}$.
It states that, in the thermodynamic limit, $\braket{E_\alpha|\hat{\mathcal{O}}|E_\alpha}$ is equal to the prediction of the microcanonical ensemble within a small energy shell \cite{Jensen85,Srednicki94,Deutsch91,Rigol08}.
When the $|c_\alpha|$'s have a sharp peak around the mean energy,
the ETH justifies the microcanonical ensemble \cite{Srednicki94,Rigol08}, and hence the canonical ensemble \cite{LandauS,Mueller13,Brandao15} (see Refs. \cite{Peres84E,Rigol12,Ikeda11,Sirker14,Ikeda15} for related scenarios).

\begin{figure}[tbp]
\begin{center}
\includegraphics[width=\linewidth]{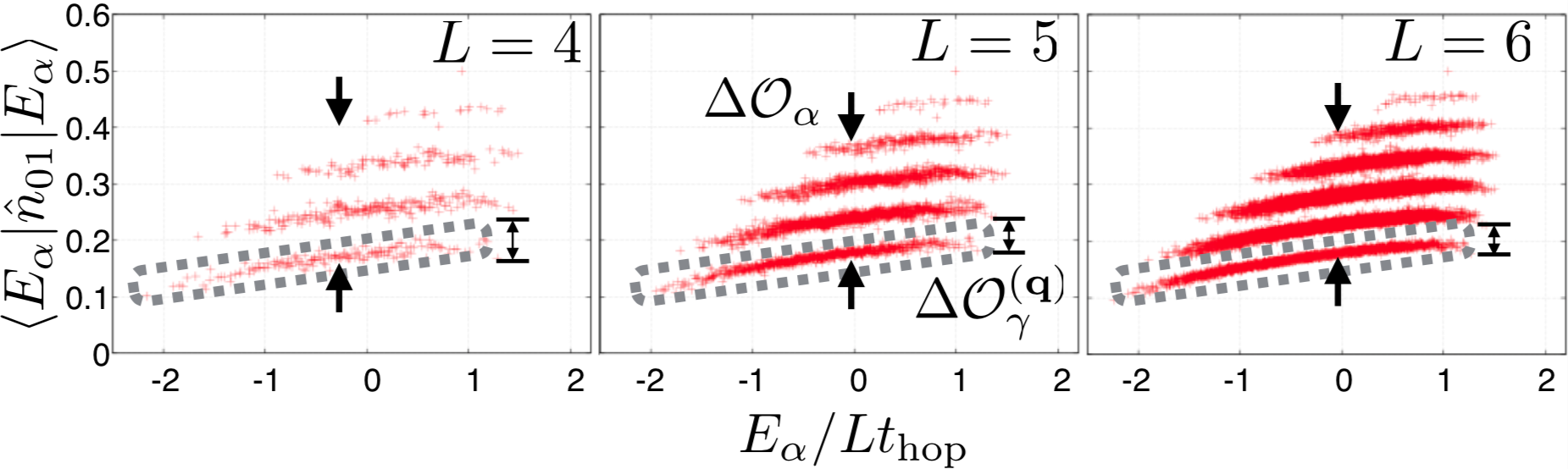}
\caption{(Color Online) The system size dependence of the EEVs for $\hat{n}_{01}$ plotted for $L=4$ (left), 5 (middle) and 6 (right). The EEV fluctuations, $\Delta \mathcal{O}_\alpha$, which are indicated by a pair of arrows, do not decrease with increasing the system size $L$. The EEVs encircled by the dotted curves shows the subset of the EEVs belonging to $\mathcal{H}_{\mathbf{q}_1}$.
The EEV fluctuations in the restricted set, $\Delta \mathcal{O}_{\gamma}^{(\mathbf{q}_1)}$, which are indicated by updown arrows, decrease with increasing $L$.}
\label{eth}
\end{center}
\end{figure}

Figure \ref{eth} shows the EEVs for $\hat{n}_{01}$, indicating the failure of the ETH when applied to the entire spectrum.
The fluctuations of EEVs (EEV fluctuations) $\Delta \mathcal{O}_\alpha$ shown by a pair of arrows in Fig. \ref{eth} do not decrease with increasing $L$.
We have found similar results for $\hat{n}_{00}$ and $\hat{n}_{11}$.

Nevertheless, the EEV fluctuations decrease if the eigenstates are restricted to each symmetry sector.
For example, each dotted curve in Fig. \ref{eth} shows the restricted eigenstates belonging to $\mathcal{H}_{\mathbf{q}_1}$.
The EEV fluctuations in this sector decrease with increasing $L$.
To be more precise, we define the EEV fluctuation $\Delta \mathcal{O}_\gamma^{(\mathbf{q})}$ in sector $\mathcal{H}_\mathbf{q}$ by
\begin{align}\label{ethfs}
\braket{E_{\gamma}^{(\mathbf{q})}|\hat{\mathcal{O}}|E_{\gamma}^{(\mathbf{q})}}=\braket{\hat{\mathcal{O}}}_\mathrm{mic}^{(\mathbf{q})}(E_{\gamma}^{(\mathbf{q})})+\Delta \mathcal{O}_{\gamma}^{(\mathbf{q})}\: .
\end{align}
Here $\ket{E_{\gamma}^{(\mathbf{q})}}$ is an energy eigenstate in $\mathcal{H}_{\mathbf{q}}$ with an eigenenergy $E_{\gamma}^{(\mathbf{q})}$, and $\gamma\:(1\leq \gamma\leq \mathrm{dim}[\mathcal{H}_{\mathbf{q}}])$ labels the eigenstate.
We also define the microcanonical ensemble in the sector $\mathcal{H}_\mathbf{q}$:
\begin{align}\label{gmic}
\braket{\hat{\mathcal{O}}}_\mathrm{mic}^{(\mathbf{q})}(E)=\frac{1}{\mathcal{N}^{(\mathbf{q})}_{E,\Delta E}}\sum_{|E-E_{\gamma}^{(\mathbf{q})}|<\Delta E}\braket{E_{\gamma}^{(\mathbf{q})}|\hat{\mathcal{O}}|E_{\gamma}^{(\mathbf{q})}}\:.
\end{align}
Here $\mathcal{N}^{(\mathbf{q})}_{E,\Delta E}$ counts the number of the energy eigenstates in $\mathcal{H}_\mathbf{q}$ within the energy shell $[E-\Delta E,E+\Delta E]$.

\begin{figure}[tbp]
\begin{center}
\includegraphics[width=7cm]{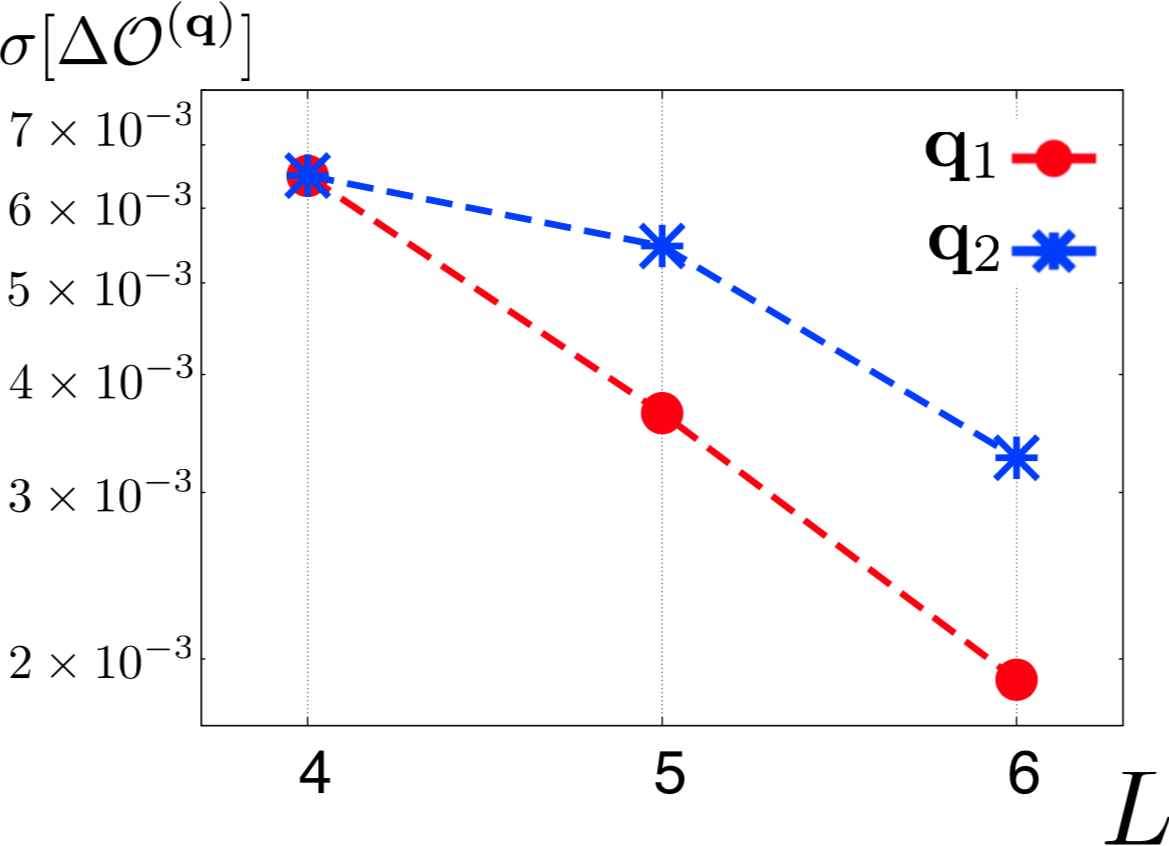}
\caption{(Color Online) The system size dependence of the standard deviation $\sigma[{\Delta {\mathcal{O}}^{(\mathbf{q})}}]$ of $\braket{E_{\gamma}^{(\mathbf{q})}|\hat{\mathcal{O}}|E_{\gamma}^{(\mathbf{q})}}-\braket{\hat{\mathcal{O}}}_\mathrm{mic}^{(\mathbf{q})}(E_{\gamma}^{(\mathbf{q})})$ within the energy shell $[E-\Delta E,E+\Delta E]$ with $\Delta E=0.18L$ and $E=0$. 
We show $\sigma[{\Delta {\mathcal{O}}^{(\mathbf{q})}}]$ for $\mathcal{H}_{\mathbf{q}_1}$ (circle) and $\mathcal{H}_{\mathbf{q}_2}$ (asterisk).
Both of them decrease with increasing $L$, indicating that the ETH holds true for each symmetry sector.}
\label{ethyuragi}
\end{center}
\end{figure}

Figure \ref{ethyuragi} shows the validity of the ETH for each sector.
We evaluate the typical magnitude of $\Delta \mathcal{O}_{\gamma}^{(\mathbf{q})}$ with $\sigma[{\Delta {\mathcal{O}}^{(\mathbf{q})}}]$. Here $\sigma[{\Delta {\mathcal{O}}^{(\mathbf{q})}}]$ is the standard deviation of $\braket{E_{\gamma}^{(\mathbf{q})}|\hat{\mathcal{O}}|E_{\gamma}^{(\mathbf{q})}}-\braket{\hat{\mathcal{O}}}_\mathrm{mic}^{(\mathbf{q})}(E_{\gamma}^{(\mathbf{q})})$ within the energy shell $[E-\Delta E,E+\Delta E]$ for $\mathbf{q} = \mathbf{q}_1 \equiv (+1,+1,\cdots,+1)$ and $\mathbf{q}_2\equiv(-1, +1, \cdots, +1)$ .
The figure shows that both $\sigma[{\Delta {\mathcal{O}}^{(\mathbf{q_1})}}]$ and $\sigma[{\Delta {\mathcal{O}}^{(\mathbf{q_2})}}]$ decrease with increasing $L$.

Assuming the ETH to be valid for each sector, the diagonal ensemble is effectively described as a statistical mixture of the microcanonical ensembles in all sectors in Eq. (\ref{gmic}), as
\begin{align}
\mathrm{Tr}[\hat{\rho}_\mathrm{d}\hat{\mathcal{O}}] 
&=\sum_\mathbf{q} p_\mathbf{q} \braket{\hat{\mathcal{O}}}_\mathrm{mic}^{(\mathbf{q})}(E_\mathbf{q})+o(1)\:. \label{gethtogge}
\end{align}
Here,
\begin{align}\label{opn}
p_\mathbf{q}=\sum_{{\gamma}} |c_{\gamma}^{(\mathbf{q})}|^2
=\braket{\psi_0|\mathcal{\hat{P}}_\mathbf{q}|\psi_0} 
\end{align}
is the occupation ratio of the sector $\mathcal{H}_\mathbf{q}$, $\mathcal{\hat{P}}_\mathbf{q}$ is the projection operator onto the sector $\mathcal{H}_\mathbf{q}$, and $c_{\gamma}^{(\mathbf{q})}\equiv \braket{E_{\gamma}^{(\mathbf{q})}|\psi_0}$. 
Also,
\begin{align}
E_\mathbf{q}=\frac{1}{p_\mathbf{q}}\sum_{\gamma} |c_{\gamma}^{(\mathbf{q})}|^2E_{\gamma}^{(\mathbf{q})}
=\frac{1}{p_\mathbf{q}}\braket{\psi_0|\mathcal{\hat{P}}_\mathbf{q}\hat{H}\mathcal{\hat{P}}_\mathbf{q}|\psi_0} 
\end{align}
is the average energy in the sector $\mathcal{H}_\mathbf{q}$.
We have assumed that $|c_{\gamma}^{(\mathbf{q})}|$'s have a sharp peak around $E_\mathbf{q}$ in deriving Eq. (\ref{gethtogge}). 
Equation (\ref{gethtogge}) depends on the $2|G|=2^{L+1}$ parameters $p_\mathbf{q}$ and $E_\mathbf{q}$.
Note that the diagonal ensemble depends on $\mathrm{dim} [\mathcal{H}]=\frac{(3L)!}{L!(2L)!}\gg 2|G|$ parameters.

We can construct the ``restricted GGE (rGGE)" with $2^{L+1}$ conserved quantities that determine $p_\mathbf{q}$ and $E_\mathbf{q}$.
If we take $\hat{Q}_0\equiv \hat{H}, \hat{Q_l}\equiv \hat{P}_l\:(1\leq l \leq L)$ and their higher-order correlations as such conserved quantities, the rGGE is constructed as
\begin{align}\label{GGGE}
\hat{\rho}_\mathrm{rGGE}
=\frac{1}{Z_\mathrm{rGGE}}e^{-\sum_{l=0}^L \kappa_l \hat{Q}_l -\sum_{l < m} \kappa_{lm} \hat{Q}_l\hat{Q}_m - \cdots}\: ,
\end{align}
where $Z_{\mathrm{rGGE}}\equiv \mathrm{tr}\{\exp[\cdots]\}$ (see Refs. \cite{Kollar08,Goldstein13,Sels14} for similar concepts).
Note that $\{\kappa_{lm\cdots}\}$ are determined from the condition
$
\braket{\psi_0|\hat{Q}_l\hat{Q}_m\cdots|\psi_0}=\mathrm{Tr}[\hat{\rho}_\mathrm{rGGE}\hat{Q}_l\hat{Q}_m \cdots ].
$
Equation (\ref{GGGE}) leads to
$\mathrm{Tr}[\hat{\rho}_\mathrm{rGGE}\mathcal{\hat{P}}_\mathbf{q}]=p_\mathbf{q}$ and $\frac{1}{p_\mathbf{q}}\mathrm{Tr}[\hat{\rho}_\mathrm{rGGE}\mathcal{\hat{P}}_\mathbf{q}\hat{H}\mathcal{\hat{P}}_\mathbf{q}]=E_\mathbf{q}$, which justifies the rGGE as the ensemble that describes a stationary state.

We conjecture that the GGE, given in Eq. (\ref{GGE}), can describe the rGGE if the supports of the observables lie in each layer.
A related conjecture made in Ref.  \cite{Fagotti13} states that we can exclude those conserved quantities that are less local than observables from the rGGE.
In our model, the products of the multiple $\hat{Q}_l$ in Eq. (\ref{GGGE}) have supports over the multiple layers.
They are thus excluded from the rGGE for $\hat{n}_{00},\hat{n}_{01}$, and $\hat{n}_{11}$, which are the sum of the local operators whose supports reside in each layer.

Before closing this section, we explain why $\overline{\delta n_\mathrm{GGE}}$ is less sensitive to $L$ for case A than for case B.
The EEV fluctuations $\Delta \mathcal{O}_\alpha$ decrease with increasing $\mathrm{dim}[\mathcal{H}]$ \cite{Ikeda15,Beugeling14}.
The restricted EEV fluctuations $\Delta \mathcal{O}_{\gamma}^{(\mathbf{q})}$ are also expected to decrease with increasing $\mathrm{dim}[\mathcal{H}_\mathbf{q}]$.
When the sectors have more negative $\mathbb{Z}_2$ parities ($q_l=-1$), they have smaller dimensions, resulting in a larger $\Delta \mathcal{O}_{\gamma}^{(\mathbf{q})}$.
Then the EEV fluctuations remain large for case A due to the sectors with negative $\mathbb{Z}_2$ parities, while they decay rapidly with increasing $L$ for case B.
Thus, $\overline{\delta n_\mathrm{GGE}}$ is less dependent on $L$ for case A than for case B.

\begin{figure}[tbp]
\begin{center}
\includegraphics[width=\linewidth]{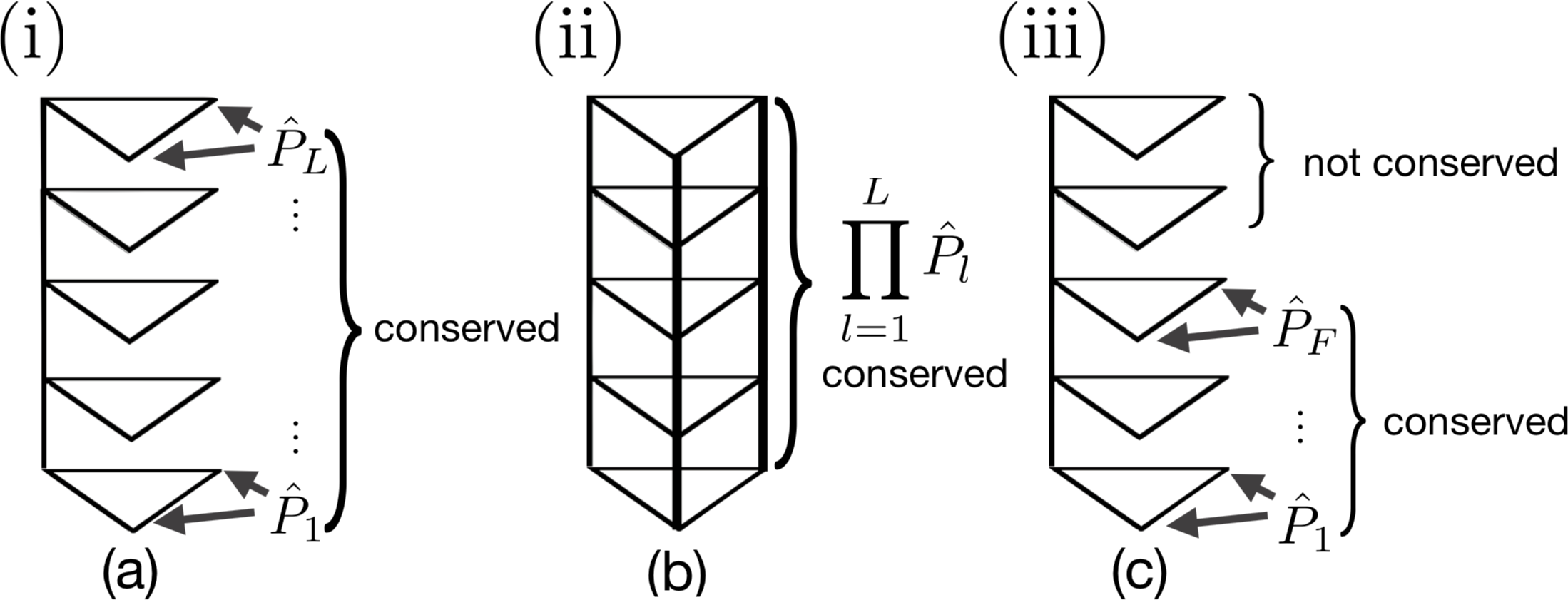}
\caption{Three models with different types of conserved quantities. (i) Model (a), which is the same as in Fig. \ref{bose}.
(ii) Model (b). Compared with (a), bosons can hop vertically between the L (or R) sites of the neighboring layers.
This model has one global $\mathbb{Z}_2$ symmetry $\prod_{l=1}^L\hat{P}_l$, instead of the local $\mathbb{Z}_2$ symmetries of model (a).
(iii) Model (c). 
This model has the local  $\mathbb{Z}_2$ symmetries only at the layers with $1\leq l\leq F$ (the case of $F=3$ is illustrated in the figure) because of randomness introduced in the other layers.
}
\label{othermodels}
\end{center}
\end{figure}

\section{Models with fewer local symmetries}

In this section, we show that the canonical ensemble works when the number of the local symmetries does not increase in proportion to $L$.
To this end, we introduce two models with fewer local $\mathbb{Z}_2$ symmetries. 

Figure \ref{othermodels} (ii) shows model (b), which has only one global $\mathbb{Z}_2$ symmetry.
The only difference from model (a) is that bosons can hop vertically between the L (or R) sites.
We assume $t_{\mathrm{L}l,\mathrm{L}(l+1)}=t_{\mathrm{R}l,\mathrm{R}(l+1)}\neq 0$,
which leads to one global conserved operator $\prod_{l=1}^L\hat{P}_l$.
This operator simultaneously swaps the sites R and L at every layer.

Figure \ref{othermodels} (iii) shows model (c), which has a fixed number $F\:(F=0,1,2,3)$ of local $\mathbb{Z}_2$ symmetries.
In this model, $t_{\mathrm{M}l,\mathrm{L}l}=t_{\mathrm{M}l,\mathrm{R}l}$ is satisfied only for $l\leq F$.
Then, it has the local $\mathbb{Z}_2$ symmetries only at the layers with $1\leq l\leq F$.
In particular, model (c) with $F=0$ has no local conserved quantity except the total energy.

\begin{figure}
\begin{center}
\includegraphics[width=\linewidth]{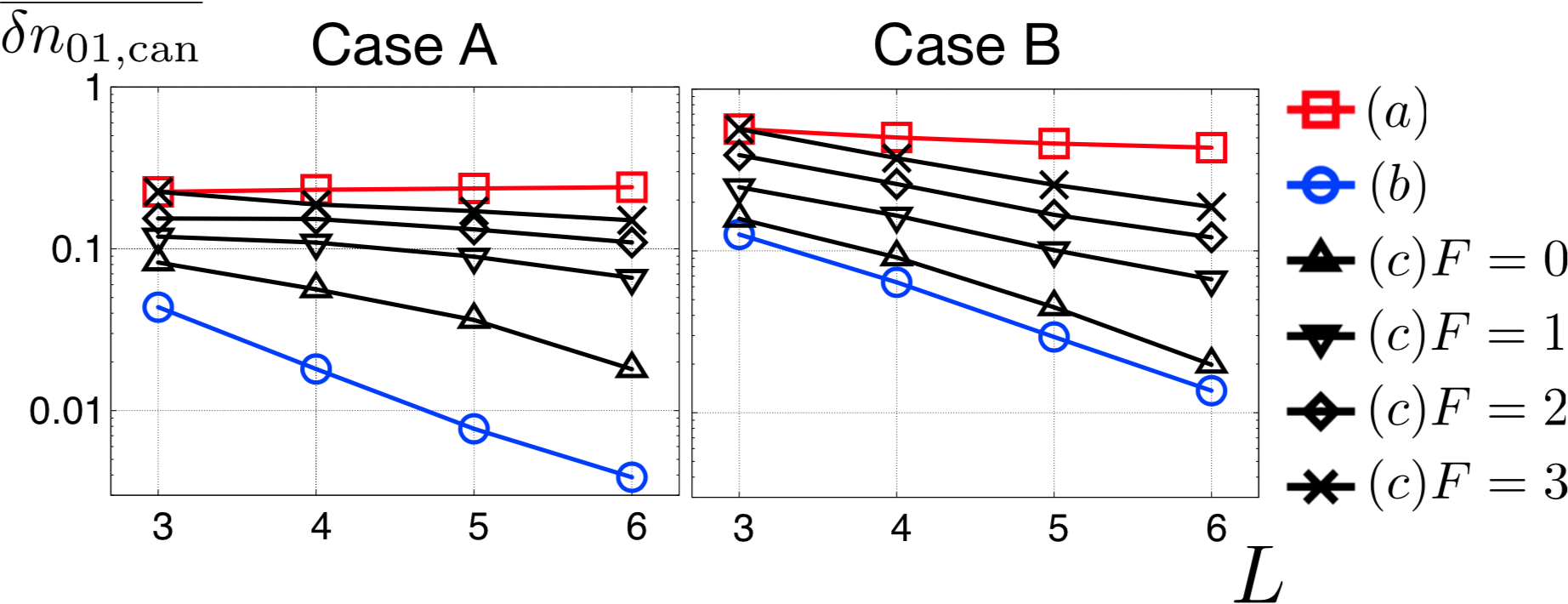}
\caption{(Color Online) Relative difference of the canonical ensemble for $\hat{n}_{01}$ compared with the diagonal ensemble in model (a) (square), model (b) (circle), and model (c) with $F=$0 (triangle), 1 (downward triangle), 2 (diamond) and 3 (cross). The left and right panels correspond to case A and case B, respectively.
For both models (b) and (c), the relative difference decreases with increasing $L$, albeit slowly for (c) with $F\geq 1$.}
\label{bc0can}
\end{center}
\end{figure}

Figure \ref{bc0can} demonstrates the validity of the canonical ensemble in the models (b) and (c), by showing the system-size dependence of $\overline{\delta n_{01,\mathrm{can}}}$.
First, in the models (b) and (c) with $F=0$, $\overline{\delta n_{01,\mathrm{can}}}$ rapidly decreases with increasing $L$, down to about one-tenth at $L=6$, compared with (a).
These results justify the canonical ensemble in these models.
Second, in the models (c), the $L$-dependence is much less sensitive for $F\geq 1$ than $F=0$.
Nevertheless, $\overline{\delta n_{01,\mathrm{can}}}$ decreases even in $F=3$, which again justifies the canonical ensemble.
We have obtained similar results for $\overline{\delta n_{00,\mathrm{can}}}$ and $\overline{\delta n_{11,\mathrm{can}}}$.
We attribute these results to the ETH, which holds less for larger $F\geq 1$ (see Appendix \ref{gethom}).

Figure \ref{waruyo} shows the $F$-dependence of $\overline{\delta n_{\mathrm{can}}}$ with $L=6$,
which shows that the canonical ensemble works better as $F$ (or equivalently, $F/L$) decreases.
This result indicates that the stationary state can be described by the canonical ensemble if the number of symmetries are much less than the system size.

\begin{figure}[tbp]
\begin{center}
\includegraphics[width=\linewidth]{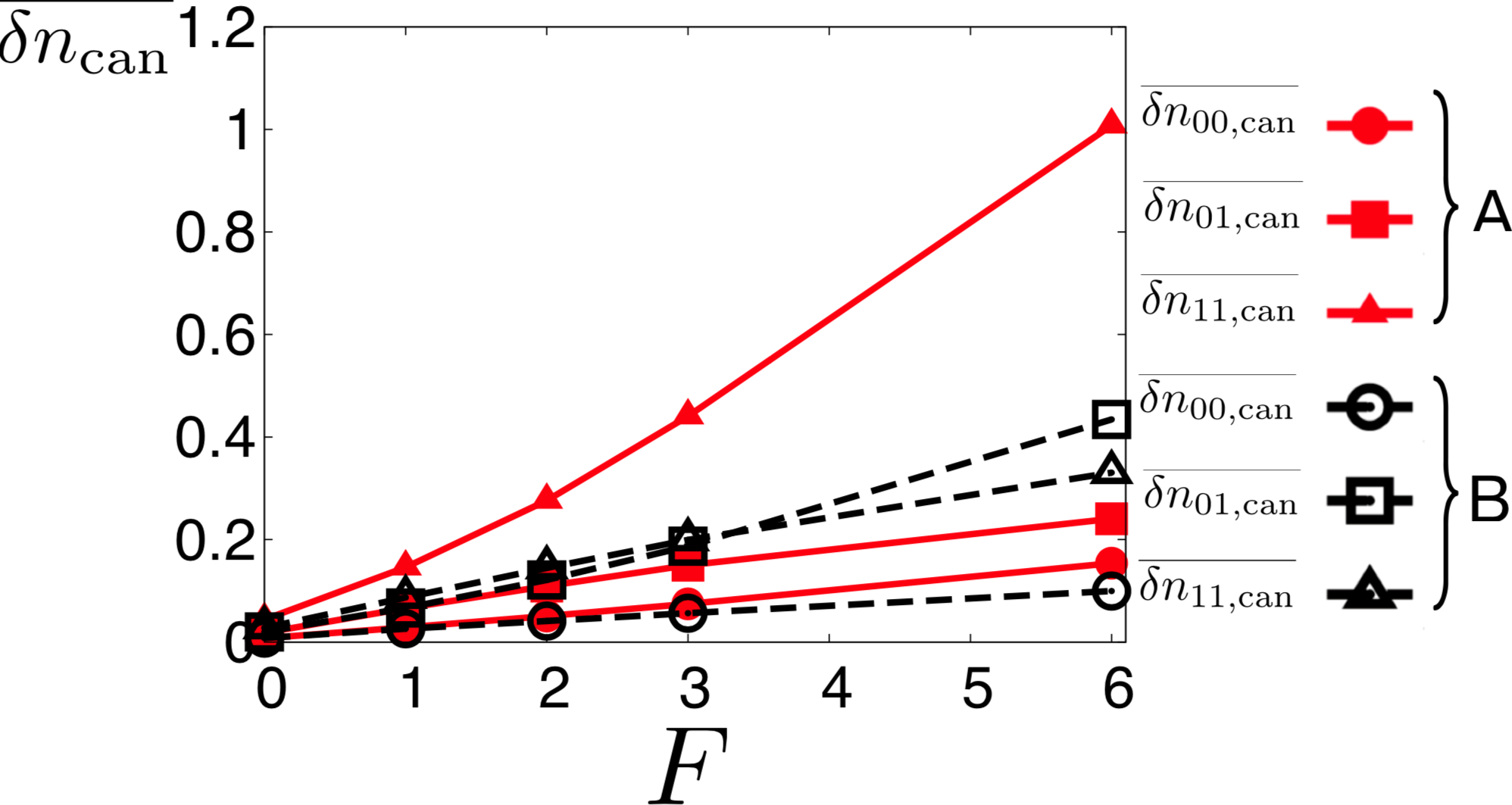}
\caption{(Color Online) $F$-dependence of $\overline{\delta n_{\mathrm{can}}}$ for $\hat{n}_{00}\text{ (circle)},\hat{n}_{01}\text{ (square)}$ and $\hat{n}_{11}\text{ (triangle)}$ with $L=6$. 
Here $F=0,1,2,3$ show the results for the models (c), and $F=6$ shows the results for model (a).
Both case A (filled) and case B (open) are shown.
The relative deviation from the canonical ensemble decreases with decreasing the value of $F$.}
\label{waruyo}
\end{center}
\end{figure}

\section{conclusions and discussions}
We have shown that stationary states for the nonintegrable model with an extensive number of local $\mathbb{Z}_2$ symmetries (Fig. \ref{bose}) can be described by the GGE rather than the canonical ensemble.
We find that the ETH holds true within each symmetry sector, but not for the entire spectrum.
We argue that this justifies the GGE if we disregard correlations among local conserved quantities.
By studying the models with only one global $\mathbb{Z}_2$ symmetry or a system-size independent number of local $\mathbb{Z}_2$ symmetries,
we find that the canonical ensemble works in these models.
Our results have clarified that we need the GGE to describe stationary states when an extensive number of local symmetries exist, even if they do not label each eigenstate. 

We still discuss some problems about the relation between the number of conserved quantities and the stationary state.
Our model (a) has an extensive number of the {\it most} local conserved quantities $\hat{P}_l$, which construct the GGE to describe the observables defined in each layer.
On the other hand, {\it in total}, this model has more than extensive number of local conserved quantities $\hat{P}_l\hat{P}_m\cdots$, which may affect the less local observables.
It is thus an open question how far we can truncate the rGGE to describe the expectation values of given observables in stationary states.
Another problem is to clarify how many symmetries are enough to create stationary states that cannot be described by the canonical ensemble.
In other words, what stationary state emerges when the number of local symmetries increases in a subextensive manner?
Since $L$ grows much faster than the number of local symmetries in this case, it is beyond the reach of our method at  present.
We leave these questions for future investigation.

\section{Acknowledgements}
We are grateful to M. Cramer, T. Deguchi, F. H. L. Essler, H. Katsura, H. Kim, M. Rigol, H. Tasaki and Y. Watanabe for fruitful discussions.
We also thank for help of S. Furukawa and S. Higashikawa at the early stage of our work. 
This work was supported by
KAKENHI Grant No. 26287088 from the Japan Society for the Promotion of Science, 
a Grant-in-Aid for Scientific Research on Innovative Areas ``Topological Materials Science" (KAKENHI Grant No. 15H05855), 
the Photon Frontier Network Program from MEXT of Japan,
and the Mitsubishi Foundation.
R. H. was supported by the Japan Society for the Promotion of Science through Program for Leading Graduate Schools (ALPS).
T. N. I. acknowledges the JSPS for the postdoctoral fellowship for research abroad.

\appendix
\section{Level statistics}\label{level}

\begin{figure}[tbp]
\begin{center}
\includegraphics[width=\linewidth]{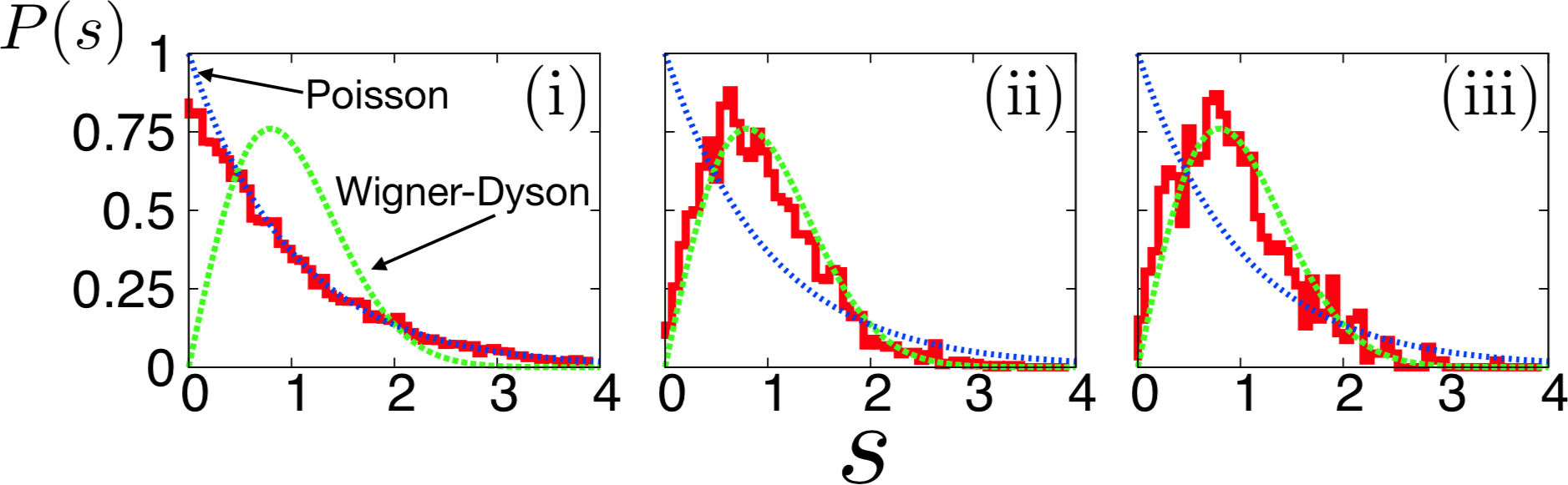}
\caption{(Color Online) (solid) Energy level statistics of the model (a) calculated for (i) the entire spectrum, (ii) the symmetry sector $\mathcal{H}_{\mathbf{q}_1}$, and (iii) the symmetry sector $\mathcal{H}_{\mathbf{q}_2}$. The Poisson ($e^{-s}$) and Wigner-Dyson ($\frac{\pi}{2}se^{-\frac{\pi}{4}s^2}$) distributions are also shown for comparison.
}
\label{strlevelall}
\end{center}
\end{figure}

We show the level statistics \cite{Haake10} of the eigenenergy spacings in the model (a) shown in Fig. \ref{othermodels}.
In nonintegrable systems that conserve only the total energy, the level statistics are expected to obey the Wigner-Dyson statistics, $P_\mathrm{WD}(s)=\frac{\pi}{2}se^{-\frac{\pi}{4}s^2}$ \cite{Bohigas84,Santos10a}.
Here $s$ is an energy-level spacing whose average is normalized to unity.
Note that we use the Gaussian orthogonal ensemble (GOE), since the Hamiltonian in Eq. (\ref{Ham}) has a time-reversal symmetry.
On the other hand, nontrivial conserved quantities lead to statistics without level repulsions, such as the Poisson statistics $P_\mathrm{P}(s)=e^{-s}$ \cite{Berry77,Santos10a}.

Figure \ref{strlevelall} (i) shows the level statistics for the entire spectrum in (a). They are closer to the Poisson statistics than the Wigner-Dyson statistics.
This reflects the existence of $\mathbb{Z}_2$ symmetries of the model \cite{Santos10b}.

Figures \ref{strlevelall} (ii) and \ref{strlevelall} (iii) show the level statistics of the eigenstates that are restricted to the sectors with $\mathbf{q}_1$ and $\mathbf{q}_2$, respectively.
They obey the Wigner-Dyson statistics.
We also find the Wigner-Dyson statistics for the other models only after specifying the symmetry sector.

\section{Occupation ratio of each symmetry sector}\label{proj}
We calculate the occupation ratio $p_{\mathbf{q}}$ defined in Eq. (\ref{opn}) for the case of the 1/3-filling.
We begin with the identity
\begin{align}
\ket{\psi}=\sum_{q_1,\cdots,q_l=\pm 1}\left\{\left[\frac{1}{2^L}\prod_{l=1}^L(1+q_{l}\hat{P}_l)\right]\ket{\psi}\right\}\:.
\end{align}
We note that $\{\cdots\}$ in this equation is a simultaneous eigenstate of the operators $(\hat{P}_1, \cdots,\hat{P}_L)$ with the eigenvalues $(q_1,\cdots,q_L)$.
Thus, the normalized projection operator is given by $\hat{\mathcal{P}}_{\mathbf{q}}=\frac{1}{2^L}\prod_{l=1}^L(1+q_{l}\hat{P}_l)$.
Since $\hat{P_l}$ swaps two sites on the layer $l$, we obtain $\braket{\psi_0^A|\hat{{P}}_{l_1}\hat{{P}}_{l_2}\cdots|\psi_0^A}=0$ and $\braket{\psi_0^B|\hat{{P}}_{l_1}\hat{{P}}_{l_2}\cdots|\psi_0^B}=1$ in case A and case B, respectively $(l_1< l_2<\cdots)$.
Expanding $\hat{\mathcal{P}}_\mathbf{q}$ and substituting the results above gives $p_\mathbf{q}=\frac{1}{2^L}\prod_{l=1}^L(1+0)$ for case A and $p_\mathbf{q}=\frac{1}{2^L}\prod_{l=1}^L(1+q_{l})$ for case B. Then 
\begin{align}
&\text{(case A)}\:\:\:p_{\mathbf{q}}=\frac{1}{2^L}\:\:\:[\text{for all }\mathbf{q}=(q_1,\cdots,q_L)]\:, \\
&\text{(case B)}\:\:\:p_{\mathbf{q}}=
\begin{cases}\label{forB}
1 & [\mathbf{q}_1=(+1,\cdots,+1)]\:; \\ 
0 & (\text{otherwise)}\:.
\end{cases}
\end{align}
We note that for the case of 1/6-filling, the result for case B is the same as in Eq. (\ref{forB}).

\section{Canonical ensemble at the infinite temperature}\label{inf}
The temperature of the canonical ensemble is infinite for both of the initial states, where the temperature is calculated from the total energy 
$E_0=0$.
We solve the equation for $\beta$,
$
0=E_0=\frac{1}{Z_\mathrm{can}}\sum_\alpha E_\alpha e^{-\beta E_\alpha}\:.
$
Since the right-hand side of this equation monotonically decreases in $\beta$,
the solution is unique.
Moreover, the Hamiltonian in Eq. (\ref{Ham}) satisfies $\mathrm{Tr}[\hat{H}]=-\sum_{\braket{ij}}t_{ij}\mathrm{Tr}[\hat{b}_i^\dag\hat{b}_j+\mathrm{H.c.}]=0$.
Here we have used $\mathrm{Tr}[\hat{b}^\dag_i \hat{b}_j]= 0$ for $i\neq j$, which can be seen by evaluating the trace in the Fock basis on the sites.
Then we obtain $\mathrm{Tr}[\hat{H}]=\sum_\alpha E_\alpha=0$, which leads to $\beta=0$.
Note that the canonical ensemble at $\beta=0$ is proportional to the identity operator,
$
\hat{\rho}_\mathrm{can}=\frac{1}{D}
$, where $D\equiv \mathrm{dim}[\mathcal{H}]$.

This canonical ensemble at $\beta=0$ gives 
\begin{align}
\braket{\hat{n}_{00}}_\mathrm{can}=\braket{\hat{n}_{01}}_\mathrm{can}=\braket{\hat{n}_{11}}_\mathrm{can}=\frac{1}{4}\:.
\end{align}
For example, for ${\hat{n}_{01}}$, we have
\begin{align}
\braket{\hat{n}_{01}}_\mathrm{can} =
\frac{1}{2^2N_bD}\sum_{i,j}e^{-i\mathbf{k\cdot (r_i-r_j)}}\delta_{z_i,z_j}\mathrm{Tr}[\hat{b}^\dag_i \hat{b}_j]\:.
\end{align}
Since the trace for $i\neq j$ vanishes, the right-hand side becomes $\frac{1}{2^2N_bD}\sum_{i}\mathrm{Tr}[\hat{b}^\dag_i \hat{b}_i]$.
Then, evaluating the trace in terms of the energy eigenstates, we have
\begin{align}
\braket{\hat{n}_{01}}_\mathrm{can} =\frac{1}{2^2N_bD}\sum_{i}\sum_\alpha \braket{E_\alpha|\hat{b}^\dag_i \hat{b}_i|E_\alpha}=\frac{1}{4}\:.
\end{align}
Here we have used $\sum_i \braket{E_\alpha|\hat{b}^\dag_i \hat{b}_i|E_\alpha}=N_b$.
Similarly, we obtain $\braket{\hat{n}_{00}}_\mathrm{can}=\braket{\hat{n}_{11}}_\mathrm{can}=\frac{1}{4}$ .

\section{ETH for the models (b) and (c)}\label{gethom}
\begin{figure}[tbp]
\begin{center}
\includegraphics[width=\linewidth]{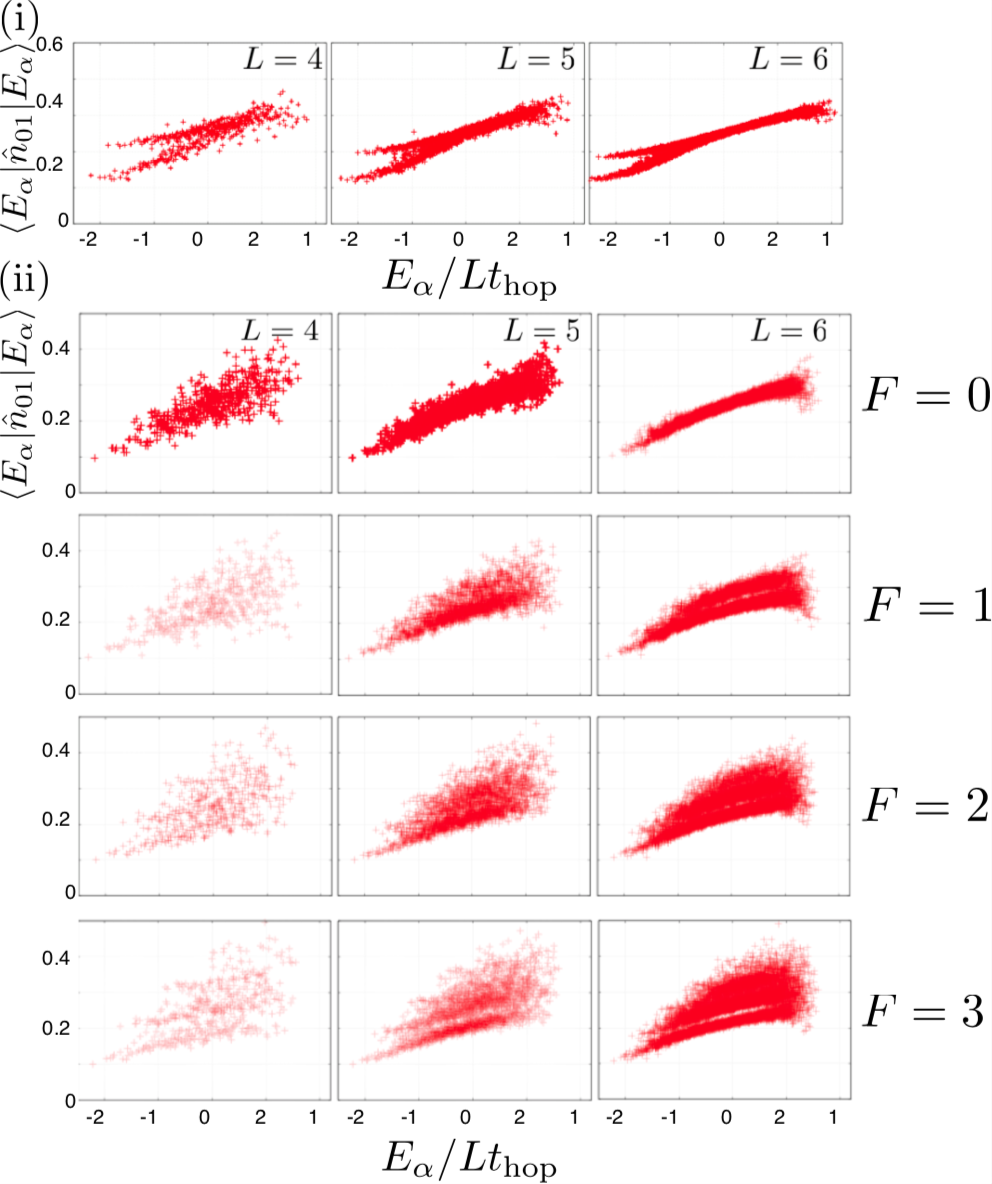}
\caption{(Color Online) (i) The EEVs for $\hat{n}_{01}$ in model (b).
The EEVs are split into two branches, but the splitting shifts toward the low-temperature region with increasing $L$.
Consequently, the EEV fluctuations decrease with increasing $L$ in the high-temperature region.
(ii) The EEVs for $\hat{n}_{01}$ in model (c).
The number of the local symmetries with $F=0,1,2,3$ increases from the top to the bottom.
The EEV fluctuations decrease with increasing $L$, but rather slowly for $F\geq 1$. 
}
\label{ethb}
\end{center}
\end{figure}

\begin{figure}[H]
\begin{center}
\includegraphics[width=\linewidth]{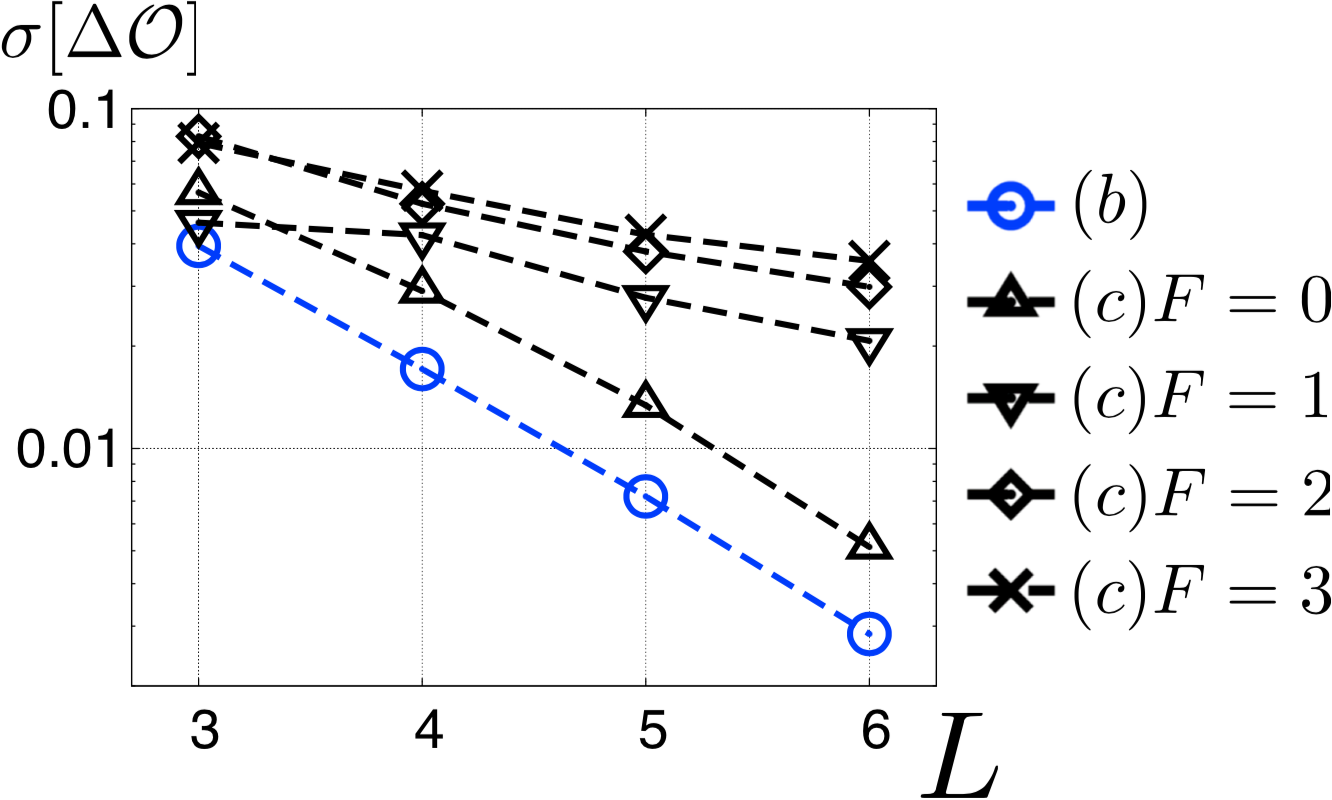}
\caption{(Color Online) The system size dependence of the standard deviation $\sigma[\Delta \mathcal{O}]$ of $\braket{E_{\alpha}|\hat{\mathcal{O}}|E_{\alpha}}-\braket{\hat{\mathcal{O}}}_\mathrm{mic}(E_{\alpha})$ within the energy shell $[E-\Delta E,E+\Delta E]$ with $\Delta E=0.18L$ and $E=0$. 
We show $\sigma[\Delta \mathcal{O}]$ in the models (b) (circle), (c) $F=0$ (upward triangle), 1 (downward triangle), 2 (diamond), and 3 (cross).
While $\sigma[\Delta \mathcal{O}]$ clearly decreases with increasing $L$, its $L$-dependence is much less sensitive for $F\geq 1$ than for $F=0$.}
\label{otheryuragi}
\end{center}
\end{figure}

Figures \ref{ethb} (i) and \ref{ethb} (ii) show the EEVs for $\hat{n}_{01}$ in the models (b) and (c), respectively.
In addition, Fig. \ref{otheryuragi} quantitatively shows $\sigma[\Delta \mathcal{O}]$, which is a typical magnitude of the EEV fluctuations $\Delta \mathcal{O}_\alpha$ around $\beta=0$.
Here $\sigma[\Delta \mathcal{O}]$ is the standard deviation of $\braket{E_{\alpha}|\hat{\mathcal{O}}|E_{\alpha}}-\braket{\hat{\mathcal{O}}}_\mathrm{mic}(E_{\alpha})$ within the energy shell $[E-\Delta E,E+\Delta E]$.
In Fig. \ref{ethb} (i), while the EEVs are split into two branches reflecting the global $\mathbb{Z}_2$ symmetry, this splitting  is shifting toward the lower temperature region with increasing $L$.
The ETH is thus expected to be true in the TDL, especially for the highly excited eigenstates (Fig. \ref{otheryuragi}).
Next, Figs. \ref{ethb} (ii) and \ref{otheryuragi} show that, although $\Delta \mathcal{O}_\alpha$ and $\sigma[\Delta \mathcal{O}]$ decrease with increasing $L$, their $L$-dependences are much weaker for $F\geq 1$ than for $F=0$.
%For $F=0$, $\Delta \mathcal{O}_\alpha$ is only attributed to the random fluctuation of the eigenstates, which decay as $\mathrm{dim}[\mathcal{H}]$.
%On the other hand, for $F\geq 1$, $\Delta \mathcal{O}_\alpha$ highly depends on the $\braket{\hat{\mathcal{O}}}_\mathrm{mic}^{(\mathbf{\tilde{q}})}(E_\gamma)-\braket{\hat{\mathcal{O}}}_\mathrm{mic}(E_\gamma)$, where $\mathbf{\tilde{q}}=(\tilde{q}_1,\cdots, \tilde{q}_F)$ is the symmetry sector.
%It can decay only by the average of the observable along the $z$ axis, with roughly $\sim L^{-1}$.
This result is consistent with the $L$-dependence of the relative difference in model (c), which is much less sensitive for $F\geq 1$ than for $F=0$.

\newpage

\bibliography{reference}
\end{document}